
%
%

\magnification\magstep 1
\vsize=8.5 true in
\hsize=6 true in
\voffset=0.5 true cm
\hoffset=1 true cm
\baselineskip=20pt

\bigskip
\bigskip
\bigskip
\vskip 5 truecm
\centerline{\bf SPIN EXCITATIONS AND SUM RULES}
\centerline{\bf  IN THE HEISENBERG ANTIFERROMAGNET}
\vskip 1.5 truecm

\centerline{S. Stringari}
\bigskip
\centerline{{\it Dipartimento di Fisica, Universit\'a di Trento,
  I-38050 Povo, Italy}}
\vskip 2 truecm

\par\noindent
{\it{\bf Abstract}. Various bounds for the  energy of  collective excitations
in the Heisenberg antiferromagnet are presented and discussed using
the formalism of sum rules.
We show that the Feynman approximation significantly overestimates
(by about 30\% in the $S={1\over2}$ square lattice) the
spin velocity due  to the non negligible contribution of multi magnons
to the energy weighted sum rule. We also discuss
a different, Goldstone type bound depending explicitly
on the order parameter (staggered magnetization).
This bound is shown to be
proportional to the dispersion of classical spin wave theory
with a q-independent normalization factor. Rigorous bounds for the excitation
energies in the anisotropic
Heisenberg model are also presented. }

\vskip 1.0truecm

\par\noindent Preprint UTF October 1993
\bigskip
\bigskip

\vskip 1.5truecm

\par\noindent
\vfill\eject

\par\noindent
{\bf 1.  INTRODUCTION}

\bigskip

In the last few years a considerable number of papers has been devoted
to the study of the Heisenberg model for antiferromagnetism,
especially in 2-D.
This interest is mainly motivated by the need for a better understanding
of the antiferromagnetic behavior of the undoped precursor insulators of the
high $T_c$ superconductors.
After the pioneering works by Anderson [1] and Oguchi [2], based on spin
wave theory, several theoretical methods have been developed
to study this problem. These range from spin-wave theory up to second order
in ${1 \over 2S}$ to series expansion methods from the Ising side
and to Monte Carlo calculations (see the review papers [3-4] for exaustive
discussions and references).

The purpose of this paper is to discuss the elementary
excitations of  the Heisenberg antiferromagnet at zero temperature
using a sum rule approach. Only recently systematic theoretical
investigations of the dispersion of spin waves in the whole
Brillouin  zone have become available [5-10]. Recent experiments [11]
in $La_2CuO_4$ with neutron scattering suggest that the dispersion follows
the predictions of classical spin wave theory with a proper renormalization
factor.
Even at low $q$, where the dispersion becomes linear,
rather relevant questions still remain to be clarified in a satisfying way.
Among them we recall the problem of
 the validity of the so called "Feynman" or single mode approximation
for the calculation of the spin velocity and of the role of multi
magnon excitations.
These questions, first discussed
by Hohenberg and Brinkman many years ago in 1D antiferromagnets [12],
have been recently addressed by Singh [9] in the $S=1/2$ square lattice.
In the present work we are mainly interested in the 2D case
and in general in systems with broken symmetries.

The paper is organized as follows:
in sect. 2 we discuss the Feynman approach to spin excitations and we prove
that it cannot reproduce the correct dispersion of spin waves at
low $q$ because of the presence of multi-magnon excitations which affect
the energy weighted sum rule also in the long wave length limit.
In sect.3 we discuss a  different bound for the energy of
elementary excitations. This bound, first introduced by Wagner [13]
many years ago,
has the form of  a Goldstone theorem and depends explicitly on the
order parameter. It can be easily calculated through the whole
Brillouin zone and in particular it exhibits the same dependence
on $q$ as the one given
by classical
spin wave theory (SWT), with a proper renormalization factor.
In sect. 4 we present results for the
anysotropic Heisenberg model. In particular we derive rigorous upper bounds
for the mass
gap in the easy-axis antiferromagnet and for the gapless dispersion law
in the case of the easy-plane antiferromagnet.

\bigskip

\par\noindent
{\bf 2. THE FEYNMAN APPROXIMATION}
\bigskip

In the following we investigate spin excitations in the framework of
the Heisenberg model for anti-ferromagnetism (AFM) characterized by the
Hamiltonian
$$
H = J \sum_{<ij>}[s^z_i s_j^z + \lambda(s^x_i s_j^x +
s^y_i s_j^y)]
\eqno(1)
$$
where $<ij>$ denotes a sum over all nearest-neighbor pairs and
$J>0$. The limits
$\lambda=0$ and $\lambda=1$ correspond to the most famous Ising and isotropic
Heisenberg models respectively. At zero temperature the
isotropic Heisenberg model is
believed to give rise to spontaneous sublattice magnetizazion also in
2D (square lattice), though quantum fluctuations
have a crucial role in reducing the value of the order parameter (actually
the $S={1\over2}$ Heisenberg model has been rigorously proven to give
rise to spontaneous magnetization only in 3D [14]).
In sections 2. and 3. we mainly discuss the isotropic case ($\lambda = 1$)
and we assume the staggered magnetization to be
oriented along the z-axis. This is  also the case for the anisotropic case
if $\lambda<1$. Conversely when $\lambda > 1$ (see sect. 4.) the axis of
(spontaneous) magnetization lies in the $x-y$ plane (easy plane).

In the following we will mainly
consider excitations generated by the spin operator:
$$
s^x_{\bf q} = {1\over \sqrt N}\sum_i s^x_i e^{i {\bf q}\cdot {\bf r}_i}
\eqno(2)
$$
These excitations are transverse with respect to the
$z$-staggered magnetization
axis. The most important among such excitations are  spin waves (magnons)
that represent the elementary excitations of the system.
Rigorous upper bounds for the energy of these excitations can be
obtained at zero temperature using the sum rule method.

The most popular
bound is  given by the Bijl-Feynman ansatz, analog of the most famous
approach employed to investigate the propagation of
density excitations in Bose superfluids [15]. It is obtained by
applying the spin operator (2) to the ground state of the system:
One finds:
$$
\mid F> = {1\over \sqrt{S^{\perp}(q)}} s^x_{\bf q}\mid 0>
\eqno(3)
$$
In eq.(3) $S^{\perp}(q) = <0\mid s^x_{-{\bf q}},s^x_{\bf q}\mid 0>$
is the transverse structure factor entering here as a normalization factor.
The excitation  energy of the "Feynman"
state is given by
$$
\epsilon_F(q) = <F\mid H \mid F> - <0\mid H\mid 0> =
{1 \over 2}{<0\mid [s^x_{-{\bf q}},[H,s^x_{\bf q}]]\mid 0>
\over S^{\perp}(q)}
\eqno(4)
$$
and provides, at zero temperature, a rigorous upper bound for the
energy $\epsilon(q)$ of the lowest state excited by the operator
$s^x_{\bf q}$. This can be directly shown by identifying the numerator and
the denominator of eq.(4) as the energy-weighted and non energy-weighted
moments of the transverse dynamic structure factor
$S^{\perp}(q,\omega) = \sum_n \mid<0 \mid s^x_{\bf q} \mid n>\mid^2 \delta
(\omega - \omega_{n0})$. In fact, using the completeness relation,
one can write
$$
\int S^{\perp}(q,\omega)\omega d\omega =
\sum_n \mid<0 \mid s^x_{-{\bf q}} \mid n>\mid^2 \omega_{n0} =
{1 \over 2}<0\mid [s^x_{-{\bf q}},[H,s^x_{\bf q}]]\mid 0>
\eqno(5)
$$
and
$$
\int S^{\perp}(q,\omega) d\omega =
\sum_n \mid<0 \mid s^x_{-{\bf q}} \mid n>\mid^2 =
<0\mid s^x_{-{\bf q}}s^x_{\bf q}\mid 0> =
S^{\perp}(q)
\eqno(6)
$$
Note that at $T=0$ the dynamic structure factor $S^{\perp}(q,\omega)$
vanishes for $\omega < 0$.

The Feynman energy (4) has been already used by several authors
to study the energy of
elementary excitations
in the Heisenberg model [12,5-6,9]. The numerator of eq.(4) can be easily
calculated
employing the commutation rules for the spin operators.
The result is
$$
{1\over 2} <0\mid [s^x_{- {\bf q}},[H,s^x_{\bf q}]]\mid 0> =
 z[f_z(1-\lambda\gamma_{\bf q}) + f_y(\lambda-\gamma_{\bf q})]
\eqno(7a)
$$
Analogously, for the $s^y_{\bf q}$ and $s^z_{\bf q}$ operators  one finds:
$$
{1\over 2} <0\mid [s^y_{- {\bf q}},[H,s^y_{\bf q}]]\mid 0> =
 z[f_z(1-\lambda\gamma_{\bf q}) + f_x(\lambda-\gamma_{\bf q})]
\eqno(7b)
$$
$$
{1\over 2} <0\mid [s^z_{- {\bf q}},[H,s^z_{\bf q}]]\mid 0> =
\lambda z(f_x+ f_y)(1-\gamma_{\bf q})
\eqno(7c)
$$
where $z$ is the number of nearest neighbors,
$\gamma_{\bf q} = {1\over z} \sum_{\bf \delta}cos{\bf q}\cdot{\bf \delta}$
and we have introduced the quantities
$$
f_x = -{J\over 2} <s^x_is^x_{i+ {\bf \delta}}>
$$
$$
f_y = -{J\over 2}<s^y_is^y_{i+ {\bf \delta}}>
\eqno(8)
$$
$$
f_z = -{J\over 2} <s^z_is^z_{i+ {\bf \delta}}> \, .
$$
Here ${\bf \delta}$ is  the lattice vector
connecting nearest neighbors. In the square lattice one has
 $\gamma_{\bf q} =
{1\over 2} (cosq_x+cosq_y)$ while in the cubic lattice
$\gamma_{\bf q} = {1\over 3}(cosq_x+cosq_y+cosq_z)$, having
set the lattice parameter equal to $1$. It is worth noticing that the
form of the energy weighted sum rule relative to  $s^z_{\bf q}$
differs from the one relative to $s^x_{\bf q}$ and $s^y_{\bf q}$. This
follows from the fact that the Heisenberg Hamiltonian (1) is invariant for
spin rotation in the $x-y$ plane.

In the isotropic case ($\lambda=1$) eq.(7a) becomes
$$
{1\over 2} <0\mid [s^x_{-{\bf q}},[H,s^x_{\bf q}]]\mid 0> =
z (f_z+f_y)(1-\gamma_{\bf q}) \, .
\eqno (9)
$$
Note that even in the isotropic limit $\lambda=1$
the quantity
$f_z$ differs from $f_y (=f_x)$ if there is spontaneous magnetization along the
$z$-axis.

At small $q$ the energy weighted sum rule (9) becomes (we consider here for
simplicity the square and cubic lattices where $\gamma_{\bf q} =
1+ {1\over z}q^2 + 0(q^2)$)
):
$$
{1\over 2} <0\mid [s^x_{-{\bf q}},[H,s^x_{\bf q}]]\mid 0> =
(f_z+f_y)q^2
\eqno(10)
$$
and exhibits the typical $q^2$ dependence characterizing
the most famous $f$-sum rule for density excitations [16].

The denominator of eq.(4) is the Fourier transform of
the two-body transverse spin correlation function. Its behavior
is dominated, at low q, by
long range correlations associated with spin waves.
Numerical results for $S^{\perp}(q)$,
based on Monte Carlo calculations [5-6] and series expansion methods [9],
are now becoming available.

{}From a general point of view the
Feynman energy (4) is expected to provide a good estimate
for the frequency of elementary excitations in Heisenberg antiferromagnets.
This system can be in fact considered
a relatively weakly interacting many body system as compared, for example,
to other strongly interacting
quantum system such as superfluid $^4He$ where the Feynman
approximation is known to overestimate in a significant way the
the energy of lowest excitations at high momenta.

An important question is howevever to understand what happens to the Feynman
approximation in the long wave length limit dominated by the propagation
of macroscopic spin waves. While in superfluid $^4$He the Feynman
ansatz is known to reproduce exactly the phonon dispersion
(in terms of sum rules this means that both the energy
weighted and non energy weighted sum rules for the density operator
are exhausted by phonons)  the situation is different
in the spin case. In fact the
non conservation of the spin current
makes the contribution of multi-magnon excitations particularly
 important  in
the low q limit. These excitations exhaust a finite fraction
of the energy weighted sum rule (EWSR) and consequently the
Feynman energy (4) does not approach the correct dispersion law
at small q. In the following we will discuss such an effect
in a quantitative way with the help of available
microscopic calculations of the spin
stiffness coefficient.

It is convenient to write the transverse dynamic spin structure
function in the following way
$$
S^{\perp}(q, \omega) =
A(q) \delta(\omega - \omega(q)) +  S^{\perp}_{mm}(q, \omega)
\eqno(11)
$$
where we have separated the sharply peaked
single magnon contribution characterized by the dipersion
law $\omega(q)$ and strength $A(q)$, from the smooth
contribution $S^{\perp}_{mm}(q,\omega)$ arising from multi magnon excitations
($S^{\perp}_{mm}(q,\omega) = 0$ for $\omega \le \omega(q)$).

The main results for the single magnon and multi magnon contributions
to the various moments of $S^{\perp}(q,\omega)$ at small $q$ are summarized
in table 1.
The main  point is the $q^2$ dependence of the strength associated with
multi-magnon excitations. This dependence differs from the $q^4$ dependence
associated, for example,
with multi-phonon excitations in Bose superfluids.
The difference is due to the fact that the current
is conseved in Bose superfluids because of translational invariance.
In the case of spin excitations the quantity $[H,s^x_q]$,
proportional to the spin current (see eq.(15) below),
is not conserved even in the low q-limit and this implies a stronger
$q^2$ dependence for the strength associated with multi-magnon excitations.
A similar behavior is exhibited by spin excitations in normal Fermi liquids
[17].
This result implies
that multi-magnon excitations affect the energy-weighted sum rule with a
term proportional to $q^2$ [18].

The occurrence of a $q^2$ contribution  to the energy weighted
sum rule due to multi-magnon excitations is clearly
exploited by the calculation of the double commutator relative
to the "longitudinal" operator
$s^z_{\bf q} = {1\over \sqrt N}\sum_i s^z_i e^{i {\bf q}\cdot {\bf r}_i}$
(see eq.(7c)) for which we find, at low q,
$$
{1\over 2} <0\mid [s^z_{-{\bf q}},[H,s^z_{\bf q}]]\mid 0>
=  (f_x+f_y)q^2
\eqno(12)
$$
This contribution, quadratic in $q$, is entirely
fixed by multi-magnon excitations
since single magnons are not excited
by $s_{\bf q}^z$.

The low $q$ contribution to the tranverse energy weighted sum rule (5,10)
arising from single magnons is given by ${1\over2}\rho_sq^2$ where
$\rho_s$ is the spin
stiffness coefficient.
This can be easily understood by using the hydrodynamic expression for the
spin velocity [19]:
$$
c^2 = {\rho_s \over \chi^{\perp}(0)}
\eqno(13)
$$
where
$$
\chi_{\perp}(q) =
2\sum_n \mid<0 \mid s^x_{-{\bf q}} \mid n>\mid^2 {1 \over \omega_{n0}}=
2\int d\omega {S^{\perp}(q,\omega) \over \omega}
\eqno(14)
$$
is the transverse magnetic susceptibility. This sum rule is expected to
be entirely exhausted, at low $q$, by the one magnon
excitation.
If the energy weighted sum rule (5,10) were also entirely exhausted
by the one magnon mode at low $q$, then the ratio
$$
lim_{q\to 0}
{1\over q^2}{\sum_n \mid<0 \mid s^x_{-{\bf q}} \mid n>\mid^2 \omega_{n0}
\over \sum_n \mid<0 \mid s^x_{-{\bf q}} \mid n>\mid^2 /\omega_{n0}}
= {2(f_z+f_y)\over \chi^{\perp}(0)}
$$
should coincide with $c^2$. The comparison between the quantities
$2(f_z+f_y)$ and $\rho_s$ then provides a direct and quantitative
information about
the contribution of multi magnons to the energy weighted sum rule.
Both the quantities $(f_z+f_y)$
and $\rho_s$ are now available through different theoretical calculations.
All the various predictions,
based on spin wave theory to second order in ${1\over 2S}$ [7,20], series
expansion from the Ising side [21] and Monte Carlo calculations [5-6]
agree with the value $ 2(f_z+f_y)=0.25$ in the $S={1\over2}$ square lattice.
Viceversa the most recent estimates for $\rho_s$ [20-22]
predict values in the range $0.18-0.20$.
Since the non energy weighted sum rule (6), entering the denominator of the
Feynman bound (4), is expected to be exhausted by the single magnon (see
eq.(19)
below), we  then conclude that
the Feynman ansatz overestimates the spin velocity
by about about 30\%. In the $S=1$ square lattice  the overestimate
is about 10\%. In table 2 we report, for completeness,
the values of various  thermodynamic parameters
relative to the 2D Heisenberg model. These values
correspond to the predictions of spin wave theory up to ${1\over (2S)^2}$ [23]
and are rather close to the ones given by the series expansion
method from the Ising side and by Monte Carlo calculations.

It is useful to study more explicitly the role of the
spin current and its connection with the
spin stiffness coefficient and the energy weighted sum rule.
To this aim let
us start from the continuity equation for the spin density
(in the folllowing the vector $\bf q$ will be taken along the
$x$-axis):
$$
[H,s^x_q] = -2iJ{1\over \sqrt N}
\sum_{<ij>}s^z_is^y_j(e^{iqx_i}- e^{iqx_j}) \equiv qj^x_{s_x}(q)
\eqno(15)
$$
definining the component of the  spin current parallel to
$\bf q$. Equation (15) provides the following
expression for the spin current at ${\bf q}=0$.
$$
j^x_{s_x}(0) = -{J\over \sqrt N}
\sum_{i, {\bf \delta}} s^z_i s^y_{i+{\bf \delta}}  \delta_x
\eqno(16)
$$
where $\delta_x =x_i-x_j$ is the $x$ component of the vector connecting the
nearest-neighbor pair $<ji>$.

The key point is that the spin current (16) is not a conserved quantity
(it does nor commute with the Hamiltonian)
and consequently, when applied to the ground state,
it can give rise to excitations
with non-vanishing strength. Such excitations are
multi-magnon states  since spin waves with ${\bf q}=0$ cannot propagate.

Let us now calculate the static response relative to the current
$j^x_{s_x}(q)$.
Due to the equation of continuity (15), this is exactly fixed by the
energy weighted sum rule for the spin operator $s^x_{\bf q}$
$$
\chi (j^x_{s_x}(q)) =
2\sum_n \mid<0 \mid j^x_{s_x}(q)\mid n>\mid^2 {1 \over \omega_{n0}} =
$$
$$
= {2 \over q^2} \sum_n \mid<0 \mid s^x_q \mid n>\mid^2 \omega_{n0} =
2(f_z+f_y)
\eqno(17)
$$
where we have taken the low q limit (10) of the energy weighted sum rule.
Both spin waves and
multi-magnon excitations affect this quantity at low $q$.
The spin wave contribution
is fixed by the spin stiffness coefficient (see the discussion above and
table 1), while
the multi-magnon contribution can be calculated through
the static response of the ${\bf q}=0$  component (16) of the spin current
operator. In conclusion we get
$$
\rho_s = 2(f_z+f_y) - \chi (j^x_{s_x}(0))
\eqno(18)
$$
Result (18) for the spin stiffness coefficient $\rho_s$
shares important analogies with the most famous expression
$\rho_s = \rho - \rho_n$
for the superfluid density of a Bose liquid.
In eq.(18)
the quantity $2(f_z+f_y)$
plays the role of the total density $\rho$, fixed by the
model independent $f$-sum rule [16], while the quantity
$\chi (j^x_{s_x}(0))$ plays the role of the normal density $\rho_n$,
defined as the low q limit of the transverse current reponse
function [24].
Note that in the case of antiferromagnetism, where the current is not
conserved,
we can safely take the $q\to 0$ limit of the current operator
for the calculation  of the multi magnon contribution to the
static current response.

It is remarkable to point out that relation (18) was obtained in an independent
way by Singh and Huse [21] starting directly  from the definition of
the spin stiffness
as helicity modulus. The full agreement between the two formal
derivations provides further support to the theory of spin hydrodynamics
and at the same time emphasizes the role played by multi
magnon excitations. Concerning this last point it
is worth noting that in the large $S$ limit
multi magnon excitations are absent, $\chi (j^x_{s_x}(0)) = 0$ and
$\rho_s$ coincides with $2(f_z+f_y)$.
Actually, using the results of spin wave theory [23],
one can easily show that the multi magnon term $\chi (j^x_{s_x}(0))$
is second order in ${1\over 2S}$, while
the longitudinal sum rule (12), dominated by multi magnons,
is  first order in ${1 \over 2S}$. This
different behavior is likely associated with the fact that longitudinal
excitations are mainly two magnon states, while the multi magnon component
of the transverse response is dominated by three magnon states.

Another important result emerging from table 1 concerns
the low q behavior of the
transverse spin structure factor (6):
$$
S^{\perp}(q)_{q\to 0} = {1\over 2} {\rho_s \over c} q
\eqno(19)
$$
accounting for the  fluctuations associated with the propagation
of long wavelength spin waves. The coefficient of linearity has been directly
calculated by Singh [9] using the series expansion method. The resulting
estimate is in reasonable agreement with eq.(19).

It is finally useful to stress that the results discussed in this
section using the sum rule technique emphasize in an explicit way
the existence of a spontaneously broken symmetry in spin space.
Different results would be obtained if one
instead decided to work with an isotropic ground state,
as happens, for example, in a numerical simulation
in a finite system. In this case the results for the  excitation energies,
obtained through the evaluation of sum rules, would
correspond to an average between transverse and longitudinal excitations
and the information on the dispersion law of elementary modes
would be consequently poorer.

\bigskip

\par\noindent
{\bf 3. ORDER PARAMETER AND EXCITATION ENERGIES}

\bigskip

The discussion of sect.2. on the behavior of the  Feynman energy
in the low q region is based
on the analysis of  the spin structure function.
The existence of spin waves with linear dispersion
must be however assumed in order to
discuss such  a behavior and cannot be predicted
using this method, unless one exploits numerically the rather difficult
low q-regime.
For this reason it is useful to derive alternative bounds for the excitation
energies which exploit more directly the low q regime. Such bounds
can be obtained with the help of an inequality due to Bogoliubov and point out
a crucial feature characterizing antiferromagnets as well as
other systems with spontaneously broken symmetries:
the existence of an order parameter. This phenomenon is known to be at the
origin of Goldstone modes which, in the antiferromagnetic case, take the
form of spin waves with a linear dispersion at low q.
This approach was first proposed by Wagner [13] to
prove the existence of Goldstone modes in an important class of
physical systems. To our knowledge it has never been used to
investigate the full q-dependence of the excitation spectrum
of Heisenberg antiferromagnets.

The starting point is the introduction of
an upper bound for the energy $\omega(q)$
 of the lowest excitation with  wave vector $\bf q$,
in terms of the ratio between the
energy weighted and the inverse energy weighted sum rules relative
to the operator $s^x_{\bf q}$:
$$
\omega^2(q) \le {\int S^{\perp}(q,\omega) \omega d\omega
\over
\int S^{\perp}(q,\omega){1 \over \omega} d\omega}
= {<[s^x_{{-\bf q}},[H,s^x_{\bf q}]]>
\over \chi^{\perp}(q)}
\eqno(20)
$$
In eq.(20) we have made use of eq.(5) and used definition (14)
for the transverse susceptibility.

The upper bound (20), holding at zero temperature, is
stronger than the Feynman one (see eq.(4)),
being based on the inverse energy weighted
sum rule $\chi^{\perp}(q)$ rather
than on the non energy weighted sum rule $S^{\perp}(q)$.
Its determination requires
however the difficult calculation of the q-dependence $\chi^{\perp}(q)$.
In the following we will combine the bound (20) with the Bogoliubov
inequality [13,25]
for the static response relative to the operator $s^x_{\bf q}$
$$
\chi^{\perp}(q) <[s_{{\bf g}-{\bf q}}^y,[H,s_{{\bf q}-{\bf g}}^y]]> \ge
\mid<[s_{-{\bf q}}^x, s_{{\bf q}-{\bf g}}^y]>\mid^2
\eqno(21)
$$
This inequality introduces the "conjugate" operator $s_{{\bf q}-{\bf g}}^y$
where ${\bf g}$ is the antiferromagnetic vector fixed by the condition
$e^{i{\bf g}\cdot{\bf R}} = 1$ when $\bf R$ connects sites in the
same sublattice and $-1$ when it connects sites in different sublattices.

Using inequality (20) and (21) we then obtain the useful rigorous
result [26]
$$
\omega^2(q) \le {<[s^x_{-{\bf q}},[H,s^x_{\bf q}]]>
<[s_{{\bf g}-{\bf q}}^y,[H,s_{{\bf q}-{\bf g}}^y]]>
\over
\mid<[s_{-{\bf q}}^x, s_{{\bf q}-{\bf g}}^y]>\mid^2}
\eqno(22)
$$
A major advantage of inequality (22) as compared to the Feynman bound (4),
is that it involves commutators both in the numerator and denominator.
In particular the quantity
$$
<[s_{-{\bf q}}^x, s_{{\bf q}-{\bf g}}^y]> = i<{1\over N} \sum_is^z_i
e^{i{\bf g} \cdot {\bf r}_i}>  \equiv im
\eqno(23)
$$
coincides with the staggered
magnetization (assumed here along the $z$-axis),
i.e. with the order parameter of the problem, and is
independent of $q$.

The full $q$-dependence of the bound (22) is then entirely
fixed by the double commutators entering the numerator.
Such commutators have been already
calculated in sect.2 (see eq.(7)). Noting that $\gamma_{{\bf q} - {\bf g}}
= - \gamma_{\bf q}$
we find the following  result
$$
\omega(q) \le {2z(f_z+f_y)\over m} \sqrt{1-\gamma_{\bf q}^2}
={2(f_z+f_y) \over mSJ} \omega^{SW}(q)
\eqno(24)
$$
where
$\omega^{SW}(q) =
z JS \sqrt{1-\gamma_{\bf q}^2}$ is the dispersion law
of classical spin wave (SW) theory [1]  and we have used the property
$f_x=f_y$.

The following remarks are in order here:

i) The rigorous bound (24) exhibits a linear behavior in $q$ for $q\to 0$,
provided the order parameter is different from zero (Goldstone
theorem).
Furthermore this bound
is symmetric by exchange of ${\bf q}$ with ${\bf g}-
{\bf q}$ and hence predicts the vanishing of elementary excitations
also at the staggered wave vector ${\bf g}$.

ii) The q-dependence of this bound is entirely contained in the
classical law $\omega^{SW}(q)$, the coefficient of proportionality
being independent of $q$. In particular from eq.(24) we obtain
the bound
$$
c \le {2(f_z+f_y) \over SmJ} c^{SW}
= 2\sqrt{2z} {(f_z+f_y) \over m}
\eqno(25)
$$
for the spin velocity
 in terms of the quantities $(f_z+f_y)$ and $m$ ($c^{SW}=\sqrt{2z}SJ$
is the prediction of classical SW theory). Using the numerical results
of table 1 for $(f_z+f_y)$ and $m$
we find $c \le 1.6 c^{SW}$ in the $S={1\over 2}$
square lattice. The bound (25)  overestimates by $\sim$ 30 \% the value
of the spin velovity calculated through equation (13) ($c=1.2c^{SW}$).
In the $S=1$ square lattice result (25) yields  $c \le 1.2 c^{SW}$
while eq.(13) gives $c=1.1c^{SW}$.
At small $q$ the quality of the new bound is hence  similar to
the one of the Feynman approximation. From a conceptual point of view
it has the advantage of exploiting directly the low $q$ behavior with
the only assumption of the existence of a broken symmetry.
It is also interesting to remark that, using the result of second order
spin wave theory [23], the bound (25) for the spin velocity coincides with
the exact value (13) up to first order terms in ${1\over 2S}$. Deviations
from the exact value are associated with multi magnon effects (terms in
${1\over
(2S)^2}$).

The dispersion
of magnon excitations in the $S={1\over2}$ square lattice Heisenberg
model has been the object of a recent Monte Carlo calculation [5]. The
authors of ref.[8] have fitted their results
with the law $\omega(q)\sim1.2\omega^{SW}(q)$ (similar results have been very
recently found also by the authors of ref.[10]), consistently with
the value of the spin velocity obtained from eq.(13).
The upper bound (24) is then found to
overestimate the magnon dispersion
by the same amount ($\sim$ 30 \%) in the whole Brillouin zone.
In fig.1 we report the prediction of the Goldstone-type bound (24) together
with the fit to the results of ref.[8] and the predictions of the
the Feynman approximation taken from ref.[9].
It is interesting
to remark that the Feynman approximation is much more accurate
near the maximum of the dispersion curve rather than in the low $q$
region where, according to the discussion of sect.2, it
overestimates the linear dispersion by $\sim$ 30 \%.

iii) Inequality (24) becomes an identity in
the large $S$ limit ($f_z = {1\over 2} S^2$, $f_x=f_y=0$, $ m = S$)
where it coincides with the prediction classical spin wave theory [1].

The Bogoliubov inequality (21) can be used to provide directly a bound for the
transverse  susceptibility $\chi^{\perp}(q)$.  Using the relation
$\gamma_{{\bf q} - {\bf g}}
= - \gamma_{\bf q}$ one finds
$$
\chi^{\perp}(q) \ge {m^2 \over 2z(f_z+f_y) (1 + \gamma_{\bf q})}
\eqno(26)
$$

At $q=0$ eq.(26) yields
$$
\chi^{\perp}(0) \ge {m^2 \over 4z(f_z+f_y)} \, ,
\eqno(27)
$$
while near
the staggered vector ${\bf g}$ one finds the typical
divergent behavior
$$
\chi^{\perp}(\mid {\bf g} - {\bf q}\mid) \ge {m^2 \over 2(f_z+f_y)
q^2}
\eqno(28)
$$
characterizing the transverse staggered susceptibility.

Once more these inequality become
 identities if one works with spin wave theory
up to first order in ${1\over 2S}$. Deviations from the exact
results for these formulae  are the direct consequence of the role
of multi magnon excitations.

It is finally useful to complete the analysis of sect.2 concerning
the contribution
to the various sum rules given by the single magnon and multi
magnon excitations in the region of the staggered vector ${\bf g}$.
The results are reported in  table 1. We note that single magnons exhaust
the transverse structure factor and susceptibility sum rules
characterized by typical infrared divergencies.
The result for the spin structure factor near the staggered vector
can be obtained with the help of the sum rule (23)
$$
\sum_n [<0 \mid s^x_{-{\bf q}} \mid n> <n \mid s^y_{{\bf q}-{\bf g}} \mid 0>
- <0 \mid s^y_{{\bf q}-{\bf g}}  \mid n> <n \mid s^x_{-{\bf q}}  \mid 0>]
$$
$$
= <[s_{-{\bf q}}^x, s_{{\bf q}-{\bf g}}^y]> = im
\eqno(29)
$$

In fact, since the magnon matrix element
$<0 \mid s^x_{-{\bf q}} \mid n>$
behaves like $\sqrt q$ at low q (see table 1 and eq.(19)), it follows that the
sum rule (29) can be satisfied only by a divergent behavior of
the magnon matrix element
$<n \mid s^y_{{\bf q}-{\bf g}} \mid 0>$ (multi magnon excitations
give rise to higher order contributions) according
to the equation
$$
<n \mid s^y_{{\bf q}-{\bf g}} \mid 0> =
<0 \mid s^y_{{\bf q}-{\bf g}} \mid n^{\prime}> = {i\over 2} {m \over
<0 \mid s^x_{-{\bf q}} \mid n>}
\eqno(30)
$$
holding for $q\to 0$.
Here $\mid n>$ and $\mid n^{\prime}>$ are single magnon states with opposite
wave vector and we have assumed, without any loss of generality, the matrix
element
$<0 \mid s^x_{-{\bf q}} \mid n> = <n^{\prime} \mid s^x_{-{\bf q}} \mid 0>$
to be real. The magnon contribution (30)
dominates the divergent behavior of the spin structure factor near
the staggered  vector that then takes the form:
$$
S^{\perp}(\mid {\bf g} - {\bf q} \mid)_{q\to 0} =  {c m^2 \over 2 \rho_s q}
\eqno(31)
$$
The above results are consistent with the rigorous inequality [27]
$$
S^{\perp}(q) S^{\perp}(\mid {\bf g}- {\bf q}\mid) \ge {1\over 4}m^2
\eqno(32)
$$
following from the uncertainty principle and holding for any value of $q$
and for any antiferromagnetic system.
According to results (19) and (31), the
uncertainty principle inequality becomes an identity in the $q\to 0$ limit.
The coefficient of the ${1\over q}$ law (31) has been recently
calculated in the $S={1\over 2}$ square lattice by Singh [9] using using the
series expansion method from
the Ising side. His prediction turns out to be
larger  by ($\sim 20$\%) than the value predicted by eq.(31). This
discrepancy remains to be understood.

Result (31) can be used to study the quality of the Feynman energy (4) near
the staggered vector ${\bf g}$. One finds:
$$
\omega_F(\mid {\bf g} - {\bf q} \mid)_{q\to 0} =
{4z(f_z+f_y)\chi^{\perp}(0) \over m^2}cq
\eqno(33)
$$
where we have used expression (13) for the spin velocity $c$.
Result (33)
overestimate the spin velocity by $\sim$ 30 \% in the $S={1\over2}$
square lattice. The enhancent coincides with the ratio between the left and
right hand sides of inequality (27) for the transverse suscptibility
and follows from the multi magnon contribution to the energy weighted
sum rule.

\vfill\eject

\par\noindent
{\bf 4. RESULTS FOR THE ANISOTROPIC HEISENBERG MEDEL}
\bigskip

The energy weighted sum rule (7) for the Heisenberg model has an
interesting behavior at low $q$
in the anisotropic case ($\lambda\ne 1$). In fact at $q=0$ eqs.(7a) and (7b)
become:
$$
\lim_{q \to 0}{1\over 2} <0\mid [s^x_{- {\bf q}},[H,s^x_{\bf q}]]\mid 0> =
z(1-\lambda)(f_z-f_y)
\eqno(34a)
$$
and
$$
\lim_{q \to 0}{1\over 2} <0\mid [s^y_{- {\bf q}},[H,s^y_{\bf q}]]\mid 0> =
z(1-\lambda)(f_z-f_x)
\eqno(34b)
$$
Conversely the EWSR relative to $s^z_{\bf q}$ vanishes with $q$ since
the Heisenberg Hamiltonian (1) conserves
the $z$-component of the spin operator.

Note that the quantities $f_z-f_y$ and $f_z-f_x$
must be positive for $\lambda<1$ and negative
for $\lambda>1$. This is a rigorous stability criterium imposed
by the positivity of the energy weighted sum rules (34).

Result (34) can be used to
derive a rigorous upper bound for the mass gap when $\lambda<1$.
In fact in this case eq.(22) yields
$$
\omega(q=0) \le {2z\over m_z} \sqrt{(f^2_z-f^2_y)}\sqrt{1-\lambda^2}
\eqno(35)
$$
where we have explicitly specified that the magnetization is along the $z$-axis
(easy axis) and used the property $\gamma_{\bf g}=-1$.

This upper bound exhibits the typical non analytic $\sqrt{1-\lambda^2}$
behavior predicted
by SWT near $\lambda=1$.
In the $S={1\over2}$ square lattice the coefficient of
proportionality of the upper bound (35) is equal to $1.9$,
compared to the value $1.3$ obtained
in ref.[28] using the series expansion method.

Using the Bogoliubov inequality (21) it is also possible to obtain
the rigorous bound
$$
\chi^{\perp}({\bf g}) \ge {m_z^2 \over 2z(f_z-f_y)(1-\lambda)}
\eqno(36)
$$
for the transverse staggered suceptibility.

Both results (35) and (36) apply only to the case $\lambda< 1$.
It is also interesting to discuss the
behavior of the
system beyond the isotropic point $\lambda=1$ where one expects the spontaneous
magnetization to occur in the $x-y$ plane (easy plane).
In the following we assume the magnetization axis to coincide with the
$x$-axis. One can find in this case a rigorous Goldstone
type upper bound similar to eq.(24). This bound is
obtained starting from inequality
(22), by replacing the operator $s^x_{\bf q}$ with $s^z_{\bf q}$
(the replacement follows from the new direction of the
magnetization axis):
$$
\omega^2(q) \le {<[s^z_{-{\bf q}},[H,s^z_{\bf q}]]>
<[s_{{\bf g}-{\bf q}}^y,[H,s_{{\bf q}-{\bf g}}^y]]>
\over
\mid<[s_{-{\bf q}}^z, s_{{\bf q}-{\bf g}}^y]>\mid^2}
\eqno(37)
$$

Using results (7) for the corresponding double commutators and
the identity  $<[s_{-{\bf q}}^z, s_{{\bf q}-{\bf g}}^y]> =
-i<{1\over N} \sum_is^x_i
e^{i{\bf g} \cdot {\bf r}_i}>  \equiv -im_x$ (staggered magnetization along
the $x$-axis), we obtain
$$
\omega^2(q) \le {4\lambda z^2 \over m_x^2}(f_x +f_y)(1-\gamma_{\bf q})
[f_z(1+\lambda \gamma_{\bf q})+f_x(\lambda +\gamma_{\bf q})]
\eqno(38) \, .
$$
yielding a
linear dispersion for $\omega(q)$ at small $q$
(the occurrence of gapless spin excitations for the easy plane
antiferromagnet has been recently pointed out in ref.[29]).
It is worth noticing however
that, differently from eq.(24) holding in the isotropic case,
the bound (38) is not symmetric by change of ${\bf q}$
with ${\bf g}-{\bf q}$ and in particular it is not gapless at the staggered
point ${\bf g}$. This reflects the fact that this system,
characterized by an anisotropy of the Hamiltonian in the $z$-direction
and by a spontaneous staggered magnetization along the $x$-axis,
exhibits two different branches in the excitation spectrum:
one excited by the operator $s^z_{\bf q}$
and for which eqs.(37-38) provide a rigorous upper bound, and one excited
by the operator $s^y_{\bf q}$. The bound for the
second branch is easily obtained
by replacing, in eq.(37), the operator $s^z_{\bf q}$
with $s^y_{\bf q}$ and $s^y_{{\bf g}-{\bf q}}$ with $s^z_{{\bf g}-{\bf q}}$.
This corresponds to
replacing ${\bf q}$ with ${\bf g}-{\bf q}$ and hence, in eq.(38),
$\gamma_{\bf q}$ with $-\gamma_{\bf q}$. Notice that this second branch
is gapless at the staggered vector ${\bf g}$.

Equation (38) provides a rigorous upper bound for the spin velocity
holding for an arbitrary value of $\lambda$ (larger than $1$ of course):
$$
c \le {2\over m_x}\sqrt{z\lambda(1+\lambda)} \sqrt{(f_x+f_y)(f_x+f_z)}
\eqno(39)
$$
Result (39) coincides with result (26) in the $\lambda \to 1$ limit
and provides a non trivial result also in the $\lambda \to \infty$
limit (XY model).

Another interesting result can be obtained for the behavior of the derivative
of the energy with respect to the transverse coupling constant $\lambda$.
This behavior is important because it characterizes the nature of the
phase transition. The derivative can be calculated starting from the
general Feynman formula
$$
{dE(\lambda) \over d\lambda} = - z(f_x+f_y)
\eqno(40)
$$
which straightforwardly follows from the form of the Heisenberg Hamiltonian
(1) and definitions (8) for $f_x$ and $f_y$. When $\lambda \to 1^-$ one has
$f^-_x=f^-_y \ne f^-_z$, while when $\lambda \to 1^+$ one has $f^+_x=f^-_z$
and $f^+_z=f^+_y=f^-_y$. This finally yields
$$
{dE(\lambda) \over d\lambda}^- = -2zf^-_y
$$
$$
{dE(\lambda) \over d\lambda}^+ = -z(f^-_z+f^-_y) \, .
\eqno(41)
$$
Using the values for $f_z$ and $f_y$ reported in table 2 (corresponding to
spontaneous magnetization along
the $z$-axis and hence to
$f^-_z$ and $f^-_y$ respectively) we find
${dE(\lambda) \over d\lambda}^- = -0.32$
and ${dE(\lambda) \over d\lambda}^+ = -0.50$. These values are
in excellent agreement
with the results obtained in ref.[30]
through a direct  Monte Carlo calculation of the energy as a
function of the coupling
constant $\lambda$.

\vfill\eject

\par\noindent
{\bf CONCLUSIONS}
\bigskip
In the present work we have derived several new results concerning
the propagation of elementary excitations in the Heisenberg antiferromagnet.
In particular:

1) We have proven that the Feynman approximation does not yield
the correct dispersion of long wavelength spin waves,
due to the role of multi magnon excitations which contribute
to the energy weighted sum rule (EWSR)  even in the low $q$ limit.
Physically this behavior originates from the fact that the
spin current is not conserved. Actually the multi-magnon contribution
to the EWSR is fixed by the static spin current polarizability
$\chi (j^x_{s_x}(0))$ (see eq.(18)).
Due to this effect, second order in ${1\over 2S}$, the Feynman
approximation turns out to overestimate  the spin velocity in the
$S={1\over 2}$ square lattice by about 30\%.

2) We have derived (sect.3) a Goldstone-type bound for the energy of spin
excitations. This rigorous bound depends explicitly on the order
parameter (staggered magnetization) and is proportional
to the classical dispersion of spin wave theory with a $q$ independent
normalization factor. It consequently
vanishes at ${\bf q}=0$ as well as at the staggered  wave vector ${\bf q}={\bf
g}$.
This bound is shown to have an accuracy similar to the one of the
Feynman approximation.

3) We have obtained useful results also for the anisotropic case (sect.4). In
particular for the easy-axis antiferromagnet we have derived a rigorous bound
for the mass gap. Viceversa the upper bound in the easy plane antiferromagnet
is proven to be gapless in agreement with the general statement of
the Goldstone theorem. We have also explicitly
calculated the discontinuity of the
derivative of the energy with respect to the transverse coupling constant
at the isotropic point.

A more systematic investigation of the structure of elementary excitations
in the anisotropic case (including the X-Y model) will be presented
in a future paper.

\vfill\eject

\par\noindent
{\bf ACKNOWLEDGMENTS}

I thank P.J. Denteneer, J.M.J. van Leeuwen, L. Pitaevskii and R.R. Singh.
for useful discussions.

\bigskip
\bigskip

\par\noindent
{\bf FIGURE CAPTION}

Dispersion of spin excitations in the $S={1\over 2}$ square lattice
($q_x=q_y$). The long-dashed line corresponds to the
$\omega(q)=1.2\omega^{SW}(q)$ fit to the Monte Carlo results of ref.[8,10];
the squares (taken from ref.[9]) correspond the Feynman bound (4),
while the full line to the Goldstone-type bound (24).
The prediction of classical spin wave theory $\omega(q)=\omega^{SW}(q)$
is also repoted (dashed line).

\vfill\eject

\par\noindent REFERENCES

\bigskip

\item{1.} P.W. Anderson, Phys.Rev. {\bf 86}, 694 (1952);R. Kubo, Phys.Rev.
{\bf 87}, 568 (1952);

\item{2.} T. Oguchi, Phys.Rev. {\bf 117}, 117 (1960);

\item{3.} E. Manousakis, Rev. Mod. Phys. {\bf 63}, 1 (1991);

\item{4.} T. Barnes, Int. J. Mod. Phys. C {\bf 2}, 659 (1991);

\item{5.} N. Trivedi and D.M. Ceperley, Phys. Rev. B {\bf 40}, 2737 (1989);

\item{6.} Z. Liu and E. Manousakis, Phys. Rev. B {\bf 40}, 11437 (1989);

\item{7.} J. Igarashi and A. Watabe, Phys. Rev. B {\bf 43}, 13456;

\item{8.} Guanhua Chen, Hong-Quiang Ding and W.A. Goddard III, Phys. Rev. B
{\bf 46}, 2933 (1992);

\item{9.} R.R.P. Singh, Phys. Rev. B, {\bf 47}, 12337 (1993);

\item{10.} Yong-Cong Chen and Kai Xiu, preprint;

\item{11.} S.M. Hayden et al., Phys. Rev. Lett. {\bf 67}, 3622 (1991);

\item{12.} P.C. Hohenberg and W.F. Brinkman, Phys. Rev. B {\bf 10}, 128 (1974);

\item{13.} H. Wagner, Z. Physik {\bf 195}, 273 (1966);

\item{14.} F. Dyson, E.H. Lieb and B. Simon, J. Stat. Phys. {\bf 18}, 335
(1978);

\item{15.} A. Bijl, Physica {\bf 8}, 655 (1940);
R.P. Feynman, in {\it Progress in Low Temperature Physics},
vol.1, ed. by C.J. Gorter (North Holland, Amsterdam, 1955), Ch.2;

\item{16.} D. Pines and Ph. Nozieres, {\it The Theory of Quantum Liquids}
(Benjamin, New York 1966),Vol.I; Ph. Nozieres and D. Pines {\it
The Theory of Quantum Liquids} (Addison-Wesley, 1990),Vol.II;

\item{17.} C.H. Aldrich III, C. Pethick and D. Pines, Phys. Rev. Lett. {\bf
37},
845 (1976);
F. Dalfovo and S. Stringari, Phys. Rev. Lett. {\bf 63}, 532 (1989);

\item{18.} To derive the low $q$ dependence of the multi magnon
contribution to the various sum rules we have made the reasonable assumption
that the corresponding
integrals involving $S_{mm}(q,\omega)$ are not critically affected
by the low energy part of the excitation spectrum.

\item{19.} B.I. Halperin and P.C. Hohenberg, Phys. Rev. {\bf 188}, 898 (1969);

\item{20.} J. Igarashi and A. Watabe, Phys. Rev. B {\bf 44}, 7247 (1991);

\item{21.} R.R.P. Singh and D.A.Huse, Phys.Rev. B {\bf 40}, 7247 (1989);

\item{22.} U.J. Wiese and H.P. Ying, preprint 1993;

\item{23.} The predictions of spin wave theory up to ${1\over(2S)^2}$
in the square lattice are given by [7,20] :

$E=4(f_z+2f_y)=-2JS^2[1+{0.316\over 2S}+{0.025\over(2S)^2}]$;

$m=S[1-{0.394\over 2S}]$;

$\chi^{\perp}={1\over 8J}[1-{0.552\over 2S}+{0.04\over (2S)^2}]$;

$\rho_s=JS^2[1-{0.236\over 2S}-{0.05\over(2S)^2}]$;

$2(f_z+f_y)=JS^2[1-{0.236\over 2S}+{0.242\over(2S)^2}]$.

\item{24.} G. Baym, in {\it Mathematical Methods in Solid State and
Superfluid Theory}, edited by R.C.Clark and G.H.Derrick (Oliver and Boyd,
Edinburgh, 1979), p.151.

\item{25.} N.N. Bogoliubov, Phys. Abh. SU {\bf 6}, 1 (1962);

\item{26.}
Inequality (22) is an example of the general
inequality
$$
\omega_0 \le {[A^{\dagger},[H,A]]><[B^{\dagger},[H,B]]>
\over \mid<[A^{\dagger},B]>\mid^2}
$$
holding at zero temperature for an arbitrary pair of operators $A$, $B$.
Here $\omega_0$ is the energy of the lowest state excited by the operators $A$
and/or $B$. Equation (22) corresponds to the choice
$A = s^x_{\bf q}$ and $B = s^y_{{\bf q} - {\bf g}}$.

\item{27.} L. Pitaevskii and S. Stringari, J. Low Temp. Phys. {\bf 85}, (1991);

\item{28.} Zheng Weihong, J. Oitmaa and C.J. Hamer, Phys. Rev. B {\bf
43}, 8321 (1991);

\item{29.} T. Barnes, et al. Phys. Rev. B {\bf 40}, 8945 (1989);

\item{30.} T. Barnes, D. Kotchan and E.S. Swanson, Phys. Rev. B {\bf 39}, 4357
(1989).

\vfill\eject

\centerline{\bf TABLE 1}

\vskip 2truecm

\settabs\+12345678912345678912345 & 12345678912345 & 12345678912345 & \cr

\+  & magnon & multi-magnons \cr
\smallskip
\hrule
\hrule
\smallskip

\+ $\omega$                       & $cq$                 & {\it const} \cr

\+ $\mid (s^x_{\bf q})_{n0}\mid^2$ & $\rho_s q/2c$  & $q^2$         \cr

\+ $\mid (s^y_{{\bf g}-{\bf q}})_{n0}\mid^2$ & $2cm^2/\rho_s q$
& {\it const}\cr

\smallskip

\+ $\sum_{n}\mid (s^x_{\bf q})_{n0}\mid^2/\omega_{n0}$ &
$\rho_s/2c^2$ & $q^2$ \cr

\+ $\sum_{n}\mid (s^x_{\bf q})_{n0}\mid^2$ & $\rho_s q/2c$
 & $q^2$ \cr

\+ $\sum_{n}\mid (s^x_{\bf q})_{n0}\mid^2\omega_{no}$ & $\rho_s q^2/2$
& $q^2$ \cr

\+ $\sum_{n}\mid (s^y_{{\bf g}-{\bf q}})_{n0}\mid^2/\omega_{n0}$ &
$m^2/2\rho_s q^2$ & {\it const} \cr

\+ $\sum_{n}\mid (s^y_{{\bf g}-{\bf q}})_{n0}\mid^2$ &
$m^2 c/2\rho_s q$ & {\it const} \cr

\+ $\sum_{n}\mid (s^y_{{\bf g}-{\bf q}})_{n0}\mid^2 \omega_{n0}$ &
$m^2c^2/2\rho_s$ & {\it const} \cr

\smallskip
\hrule
\hrule
\vskip 2.5truecm
{\parindent0pt Matrix elements, excitation energies and sum
rule contributions from one-magnon and multi-magnon excitations at $T=0$.

\vfill\eject

\centerline{\bf TABLE 2}

\vskip 2truecm

\settabs 7\columns

\+  & $E$  & $m$ & $\chi^{\perp}$ & $\rho_s$ & $c$ & $f_z+f_y$ & $f_z-f_y$
\cr
\smallskip
\hrule
\hrule
\smallskip

\+ $S={1\over 2}$      & -0.67  & 0.30 & 0.061 & 0.18 &  1.7 & 0.125 & 0.04 \cr
\+ $S=1$               & -2.33  & 0.80 & 0.092 & 0.87 &  3.1 & 0.47 & 0.25 \cr

\smallskip
\hrule
\hrule
\vskip 2.5truecm
{\parindent0pt Parameters of the isotropic 2D AF Heisenberg model
predicted by spin
wave theory up to second order in ${1\over(2S)^2}$[23].
The Heisenberg coupling constant $J$ has been set equal to $1$ and
magnetization is taken along the $z$-axis.

\bye